\documentstyle[11pt,aaspp4]{article}
\def\ltsim{\, {}^<_\sim \,}
\def\gtsim{\, {}^>_\sim \,}
\def\etal{{\it et al.}}
\def\ie{{\it i.e.}}
\def\eg{{\it e.g.}}
\def\cf{{\it cf.}}

\slugcomment{Submitted to Astron.J.}

\lefthead{Harris}
\righthead{A Globular Cluster in NGC 5128}

\begin{document}
\def\ltsim{ \,{}^<_\sim\, }
\def\gtsim{ \,{}^>_\sim\, }
\def\etal{{\it et al.}}
\def\eg{{\it e.g.}}
\def\ie{{\it i.e.}}
\def\cf{{\it cf.}}

\title{A Color-Magnitude Diagram for a Globular Cluster\\
    in the Giant Elliptical Galaxy NGC 5128}

\author{Gretchen L.~H.~Harris and G.~B.~Poole}
\affil{Department of Physics, University of Waterloo \\
    Waterloo ON N2L 3G1, Canada; glharris,gbpoole@astro.uwaterloo.ca}

\and

\author{William E. Harris}
\affil{Department of Physics and Astronomy, McMaster University \\
    Hamilton ON L8S 4M1, Canada; harris@physics.mcmaster.ca}

\begin{abstract}
The {\it Hubble Space Telescope} has been used
to obtain WFPC2 $(V,I)$ photometry for a large sample of
stars in the outer halo of the giant elliptical NGC 5128, 
at a distance 4 Mpc beyond the Local Group.  
The globular cluster N5128-C44,
at the center of the Planetary Camera field,
is well enough resolved to permit the construction of
a color-magnitude diagram (CMD) for it
which covers the brightest two magnitudes of the giant branch.
The CMD is consistent with that of a normal old, moderately low-metallicity 
([Fe/H] $\simeq -1.3$) globular cluster,
distinctly more metal-poor than most of the field halo stars
at the same projected location (which average [Fe/H] $\sim -0.5$).  
This is the most distant globular cluster in which direct 
color-magnitude photometry has been achieved to date, and the 
first one belonging to a giant E galaxy.
\end{abstract}



\section{Introduction}

Globular clusters are typically found in much larger numbers
within giant elliptical galaxies than in large spirals such
as the Milky Way or M31, and in extreme cases such as cD galaxies
the total population can exceed 20,000 clusters per galaxy (\eg\ \cite{har91}).
It is conventionally assumed that the clusters
in these remote galaxies must individually resemble the familiar 
Milky Way old-halo globulars in their
ages, metallicities, and stellar compositions. 
But making direct tests of this fundamental assumption 
in different galaxies is a difficult task.
For even the closest large galaxy, M31, it has become possible only 
recently to study the stellar compositions of its old-halo clusters
through direct color-magnitude photometry (\eg, \cite{fus96};
\cite{rich96}).  The situation is worse for giant E galaxies,
which are all far beyond the Local
Group and in which individual clusters are seen as no more than faint,
unresolved semi-stellar objects even under the best imaging
conditions.  The only reasonably direct tests of their
nature are ones based on
integrated light -- color indices, spectral properties, line
strengths -- which verify that their 
characteristics at least roughly match those of the
Milky Way globulars (see, \eg, \cite{coh97} for a
comprehensive recent spectroscopic 
study of clusters in M87; or \cite{jab96} for
spectral characteristics of clusters in NGC 5128).
However, the information derived from integrated light 
must, by definition, be
heavily smoothed over the composite stellar population.  
A color-magnitude diagram (CMD) for
any such object, even with limited depth, would 
provide a uniquely direct test of their nature.

\section{Observations}

By far the closest giant elliptical galaxy is NGC 5128, the dominant
member of the Centaurus group
at $d \sim 4$ Mpc (the next nearest gE is NGC 3379 in the Leo group, 
which at $d \sim 10$ Mpc is a full two magnitudes more distant).
Although its innermost few kiloparsecs are
affected by the well known dust lanes and gas, the main body
of the galaxy and its extensive halo are by all indications those of
a basically normal giant elliptical (\eg, \cite{gra79}; \cite{ebn83};
\cite{hui95}; \cite{sor96}; \cite{sto97}).  Since interactions
and mergers of the type now happening in NGC 5128 
are now recognized to be rather
common occurrences in the ongoing histories of large
galaxies, it is entirely reasonable to expect that
its underlying stellar composition will be typical of gE's.  
In addition, its population of $\sim 1500$ globular clusters 
(\cite{har84}) has color and luminosity distributions that clearly resemble
the clusters found in other, more remote gE's such as in
Virgo and Fornax (see \cite{hghh92}
[hereafter HGHH], \cite{hes84}, and \cite{har88}).  NGC 5128 unquestionably
offers our best chance to investigate globular clusters in 
a giant E galaxy in the best possible depth and detail.

In HST/WFPC2 program GO-5905, executed in two visits on
1997 Aug 16 and 24, we obtained deep images of a field
in the outer halo of NGC 5128. The PC1 CCD was centered on 
the globular cluster N5128-C44, which is
located at a projected distance of $R_{gc} = 18\farcm32$ from
the center of NGC 5128, equivalent to $21.3 (d/4)$ kpc.
This location in the halo is far outside the core regions contaminated
by the dust lanes and thus should be representative of the 
old-halo population of the elliptical (see HGHH for complete wide-field
finder charts and a compilation of the relevant cluster data).
C44 itself is a luminous, moderately low-metallicity cluster according
to the photometry published in HGHH:  at an integrated magnitude
$V = 18.48$ ($M_V = -9.87$ for a distance modulus $(m-M)_V = 28.35$;
see below) it is similar to the luminous halo globulars in the 
Milky Way such as M3 or NGC 2419, while its 
$(C-T_1)$ color yields an estimated metallicity
[Fe/H] $= -1.4 \pm 0.25$, quite similar to the mean metallicity of
the Milky Way halo clusters and field stars.  Lastly, the cluster
radial velocity of 520 km s$^{-1}$ is almost identical with the
systemic mean of $\sim 540$ km s$^{-1}$ for its host galaxy (HGHH;
\cite{har88}; \cite{hes86}).

Our WFPC2 observations consisted of 10 images totalling
12800 seconds in each of the two standard filters
F606W ($V$) and F814W ($I$),
with individual exposures sub-pixel
shifted over five separate positions.  The images were 
reregistered and combined to form composite deep $(V,I)$ images
free of cosmic rays and bad pixels.
By simultaneously imaging both a metal-poor globular cluster
and the halo field stars, our
original intention was to gain the most direct possible comparison
between their two mean metallicities, as well as to give us
an additional consistency check on the
metallicity scale from our $(V-I)$ colors.
Our photometry from the WF2,3,4 images has been used to generate
a color-magnitude diagram (CMD) for the galaxy halo, covering
about 2.5 magnitudes of the old red giant branch (RGB) and
containing more than 10,000 stars.  That material is described
in a companion paper along with a more detailed outline
of the data reduction (\cite{har98}); here, we 
present the results for the PC1 field and the globular cluster.

In the following discussion, we adopt a distance to NGC 5128
of 4.0 Mpc, based on planetary nebula luminosity functions
(\cite{hui93}), surface brightness fluctuations (\cite{ton90}; 
\cite{hui93}), and the luminosity of the RGB tip (\cite{har98}; 
\cite{sor96}), all normalized to
a contemporary Local Group distance scale in which $d(M31) = 770$ kpc
(\eg, \cite{vdb95}; \cite{fer98}; and references cited there).
The foreground reddening is adopted as $E(B-V) = 0.11$ and
$E(V-I)=0.14$, giving $(m-M)_V = 28.35$.

\section{Data Reduction}

The image of the PC field around C44 is shown in Figure 1.
The cluster subtends only a small angle (at $d = 4$ Mpc,
typical globular cluster 
core and tidal radii of $r_c \sim 2$ pc, $r_t \sim 50$ pc
would have angular sizes of $0\farcs1$ and $2\farcs5$, superimposed on
the PC1 image scale of $0\farcs046$ per pixel).  Thus almost all of
its stars are severely crowded and stellar photometry
is restricted to the brighter stars on its
outskirts.  But the task is not hopeless.
An excellent analogy of what we can expect to accomplish is to
compare with the globular clusters in M31: 
in terms of spatial resolution and scale, imaging M31 {\it from 
the ground} under good seeing conditions of $\simeq 0\farcs5$
is closely comparable with imaging NGC 5128 (5 times more
distant) with the $0\farcs1-$resolution of HST.  
Successful ground-based studies of several M31
clusters were indeed achieved in the pre-HST era
(\cite{cou95}; \cite{hea88}; \cite{chr91}).  
In any one of these clusters, typically two or three dozen of the
RGB stars could be isolated and measured, with representative
photometric scatter of $\sigma(V-I) = \pm0.2-0.3$
driven by the crowding and background nonuniformity.
Although the precision of these studies 
was unavoidably low, it is important to note
that the subsequent much deeper and more precise photometry 
carried out later with HST (see, \eg,
\cite{fus96}; \cite{rich96}) verified that this pioneering ground-based
photometry was able to produce {\it systematically} accurate colors
and mean RGB loci.

\begin{figure}
\plottwo{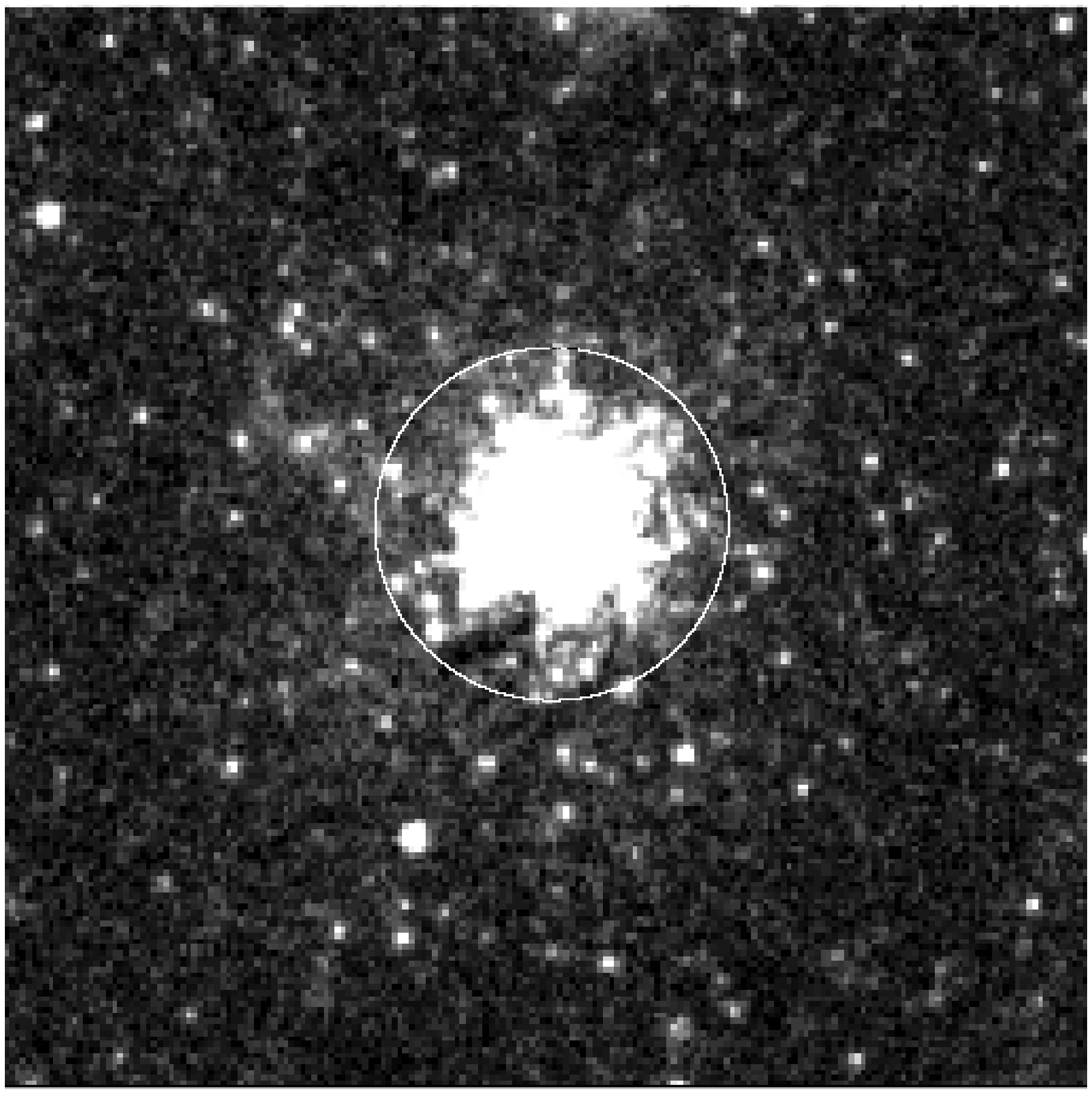}{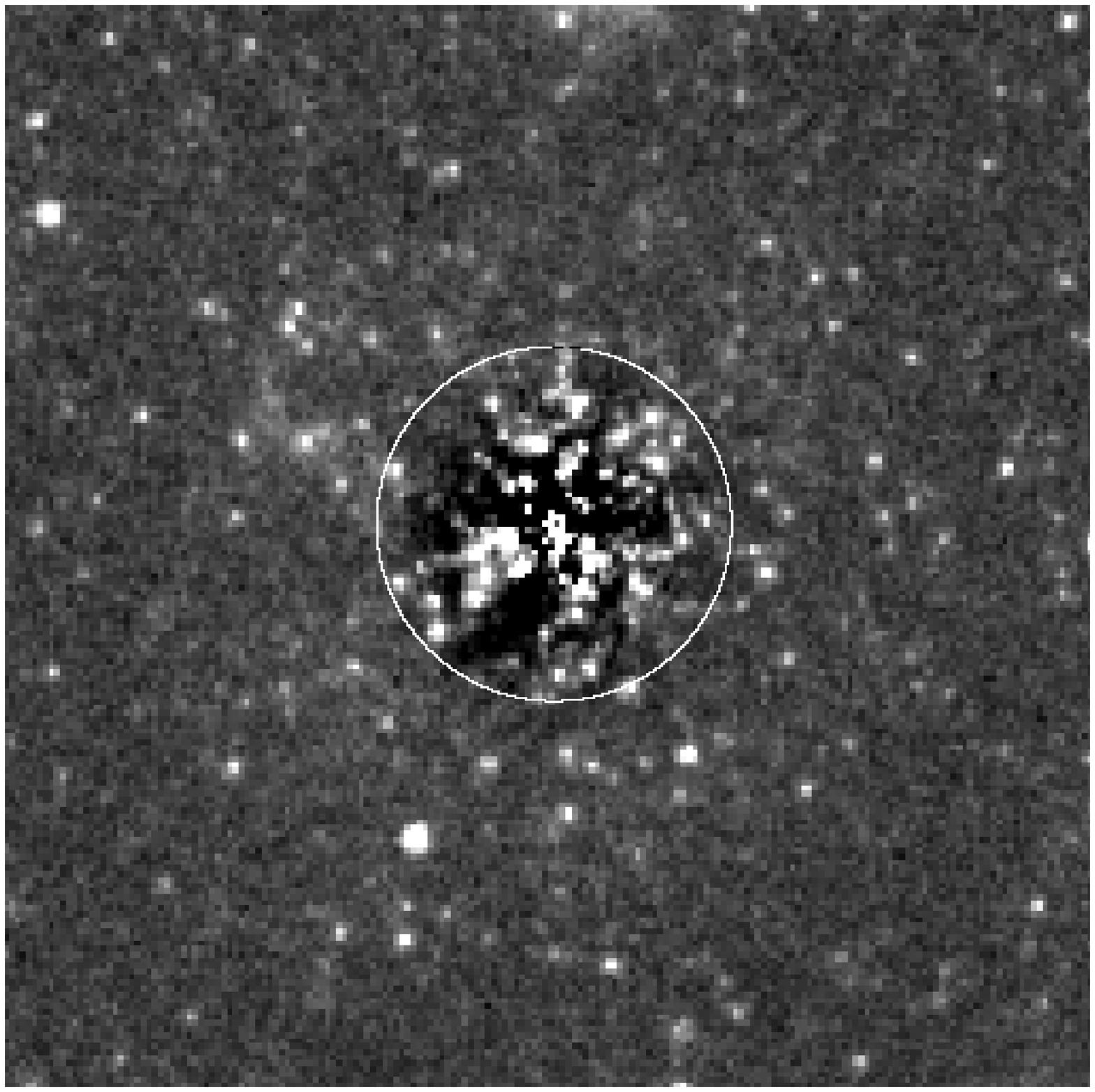}
\caption{(a) {\it First panel:} The central $190 \times 190$ pixels
($8\farcs7 \times 8\farcs7$) of the Planetary Camera WFPC2 field,
centered on globular cluster C44 in the halo of NGC 5128.
The image is a 12800-second sum of 10 exposures in $V$ (F606W).
The smooth unresolved light of the cluster has been partially
subtracted to emphasize the resolution into stars of its
outer envelope.
The inscribed circle denotes the radius (30 px $= 1\farcs4$) of the 
inner zone used to define the cluster CMD.
(b) {\it Second panel:} The same PC1 field after removal of elliptical
isophotes.  \label{fig1}}
\end{figure}

We first extracted a $190 \times 190$ pixel subsection ($8\farcs74 \times
8\farcs74$) from the PC1 image, centered on C44 (Figure~\ref{fig1}).  To eliminate
as much of the unresolved light of the cluster as possible, we 
used the ellipse-fitting code within STSDAS to derive an isophotal
contour model for the cluster, and subtracted this from the raw image.
The result (second panel of Figure~\ref{fig1}) 
was then used for stellar photometry through
the DAOPHOT II code (\cite{ste92}).  Three iterations of
the usual sequence FIND/PHOT/ALLSTAR were employed, along with
a very small psf-fitting radius (0.8 fwhm), to find and extract
the individual stars near the cluster.
As a check on this procedure, we also tried removing the unresolved
cluster light by a first pass of finding and removing stars, smoothing
the remaining light by median filtering, subtracting that from the 
original image, and repeating the process (see \cite{cou95}).
The overall results for the stellar photometry 
were quite similar; however, the first
(ellipse-fitting) procedure allowed us to reach slightly further in
toward cluster center and to define the CMD a little more narrowly.

Zeropoints and color terms for the conversion of instrumental
magnitudes into $(V,I)$ followed the prescriptions of
\cite{hol95} and are discussed completely in G.~Harris \etal\ (1998).

\section{The Color-Magnitude Diagram}

\begin{figure}
\plotone{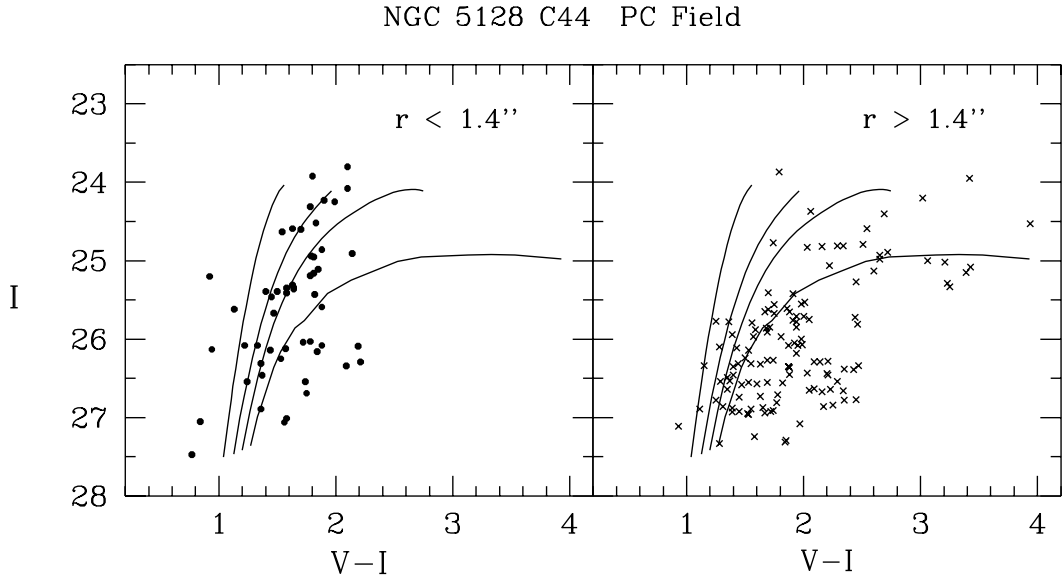}
\caption{Color-magnitude data for all measured stars
in the central PC field.  Solid dots (left panel) denote stars within
the inner 30-px ($1\farcs4$) circle; crosses (right panel)
denote objects outside that circle, which are mostly 
RGB stars in the halo of NGC 5128.
The four lines superimposed on the diagram are fiducial lines
for Milky Way globular clusters of four different metallicities
(\protect\cite{dac90}; \protect\cite{gua98});
from left to right, they are M15 ([Fe/H] $= -2.2$),
NGC 1851 ($-1.3$), 47 Tuc ($-0.7$), and NGC 6553 ($-0.25$).
The lines have been placed assuming an intrinsic distance
modulus for NGC 5128 of $(m-M)_0 = 28.0$ and reddening
$E(V-I) = 0.14$ (see text).
\label{fig2}}
\end{figure}

The resulting CMD is shown in Figure~\ref{fig2}, in which all measured data points
are plotted without any selection criteria applied. 
The stars closest to the cluster ($r < 1\farcs4 \simeq 30$ px, 
which we will call
the ``inner zone'') are systematically bluer 
than the surrounding field-halo stars,
suggesting that this particular cluster
has distinctly lower metallicity than the average of the halo stars.
To define a cleaner CMD for the cluster,
we culled the data in two additional ways:  (a) stars with very
poor goodness-of-fit values ($\chi > 4$, although the great majority
of stars had $\chi < 2$) were eliminated;
and (b) individual stars were removed from the
inner-zone CMD in proportion to their presence in the outer ($r > 30$ px)
zone.   For this latter step,
we divided the CMD into a grid of boxes of size $\Delta I \simeq 0.5$,
$\Delta(V-I) \simeq 0.2$.  Then, noting that the area of the outer zone
(70.0 arcsec$^2$) is $\sim 11$ times larger than that of the inner zone
(6.4 arcsec$^2$), we removed one star from the inner-zone CMD in any
box which also contained $\sim 10 \pm 3$ stars from the outer zone.
(This star-by-star extraction will actually remove too {\it few} stars
for $I \gtsim 25.5$, where the detection completeness is much lower
in the highly crowded cluster than in the surrounding field. Thus we
expect the fainter parts of the cluster CMD to remain partially
contaminated, but by only two or three stars at worst.)

\begin{figure}
\plotone{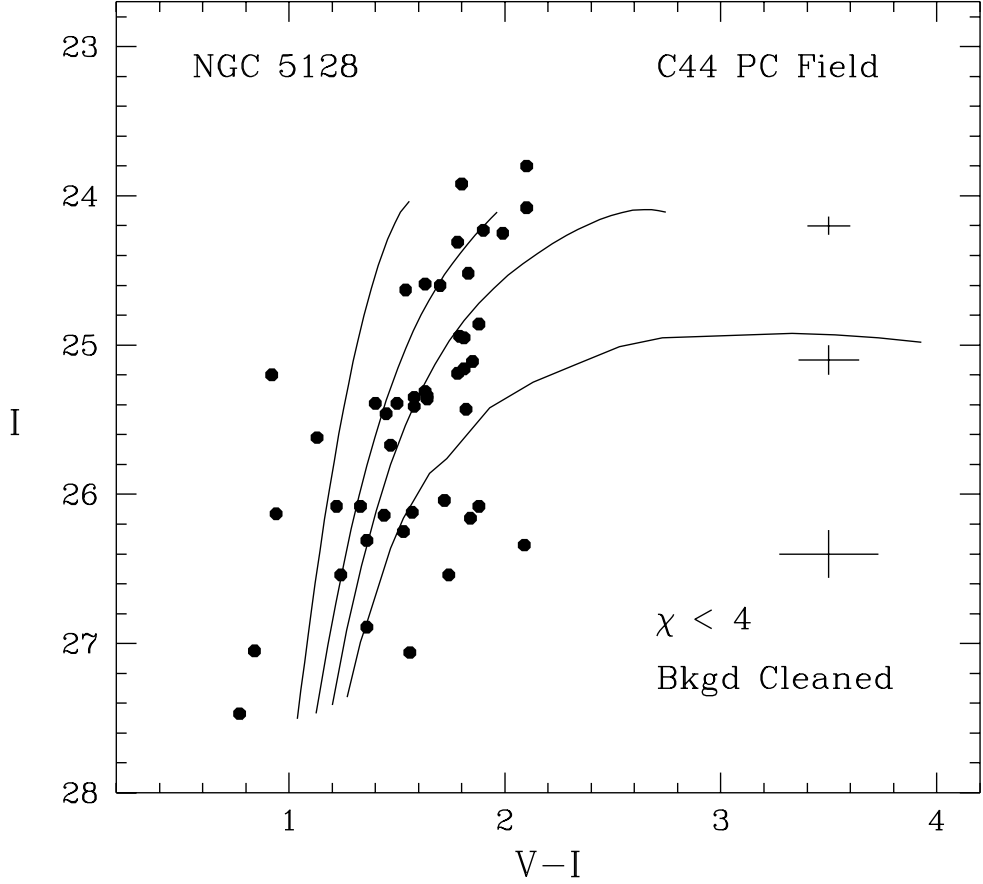}
\caption{Statistically cleaned CMD for the globular cluster C44.
Stars with poor PSF fits ($\chi > 4$) 
have been rejected, and background contamination
in proportion to the area of the inner zone has been removed
as described in the text.  The fiducial cluster sequences for
four different metallicities are the same as in Fig.~2.
Photometric measurement uncertainties from the ALLSTAR fitting
algorithms are indicated for three different magnitude levels
by the error bars at right.
C44 appears to have a moderately low metallicity [Fe/H] $\simeq -1.3$
based on the color of its RGB, and in close agreement with
the metallicity from its integrated $(C-T_1)$ color index.
\label{fig3}}
\end{figure}

The ``cleaned'' CMD for the cluster is shown in Figure~\ref{fig3}.
The 45 remaining points now scatter more closely around the fiducial
line for [Fe/H] $ = -1.3$ over the 2.5-magnitude run of the data, 
particularly at the bright RGB tip where the stellar photometry should
be the most reliable.  This metallicity estimate is uncertain
by perhaps $\simeq \pm 0.3$ dex taking into consideration both the
random photometric scatter and the uncertainties in reddening and
distance moduli; but it is
entirely consistent with the previous metallicity calibration
[Fe/H]$_{C-T_1} = -1.4$ from its integrated colors (HGHH).
The typical color uncertainty
$\sigma(V-I) \simeq \pm0.15 - 0.2$ per star, along with some residual
contamination by field stars, is enough to be responsible for
the remaining scatter we see in the CMD.

\section{Radial Profile and Structural Parameters}

\begin{figure}
\plotone{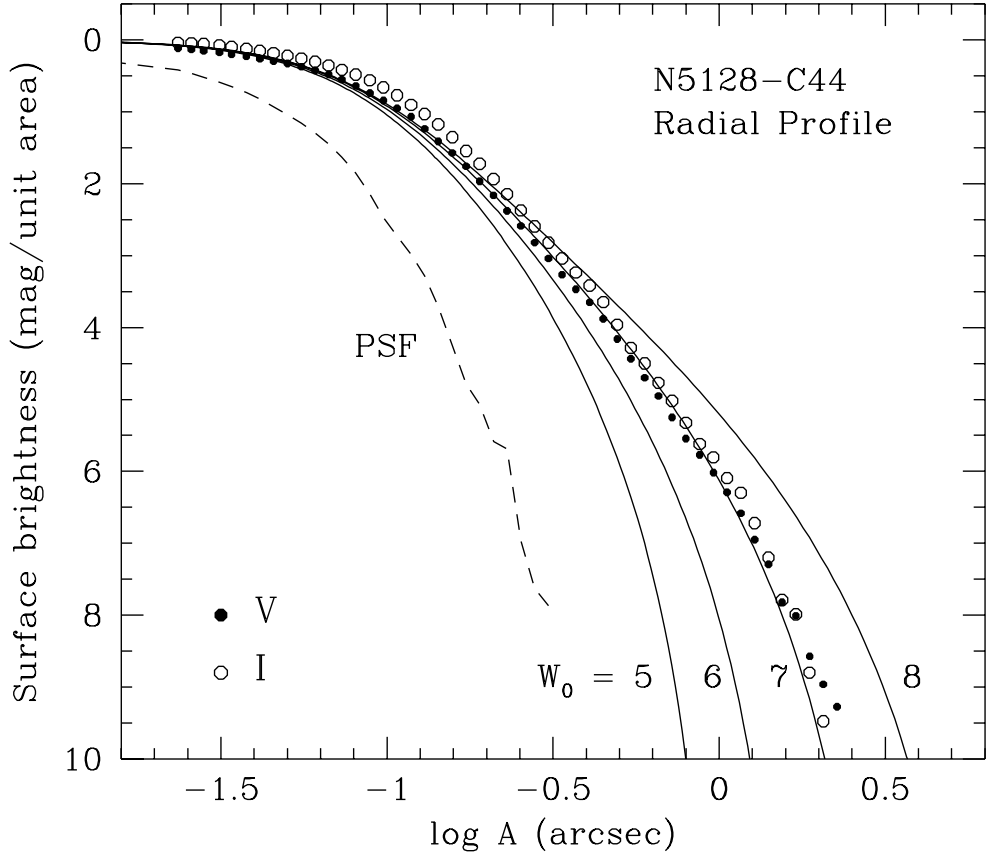}
\caption{Radial profile of the cluster light, plotted as surface
brightness (magnitudes per unit area relative to the central intensity)
against log $A$ (semi-major axis) in arcseconds.  {\it Solid dots}
denote the surface intensity in $V$ of the smoothed isophotal
model of the cluster; {\it open circles} denote the intensity profile 
in $I$.  The dashed line at left shows the profile of the stellar point-spread
function.  {\it Solid lines} superimposed on the data points are
profiles for four King models with different values of the central
potential parameter $W_0$ (curves of increasing central concentration 
go from left to right).  We estimate the cluster to have a core radius 
$r_c \simeq 0\farcs08$ (about twice as large as the PSF core)
and tidal radius $r_t \simeq 3''$.
\label{fig4}}
\end{figure}

\begin{figure}
\plotone{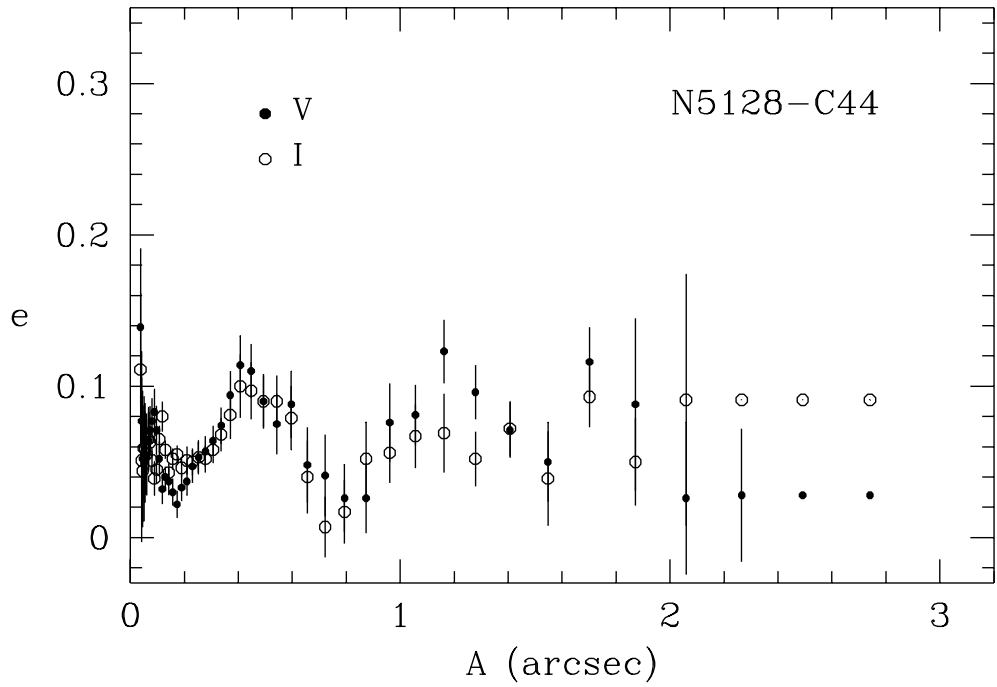}
\caption{Eccentricity $e$ of the cluster profile, as a function of
semi-major axis $A$.  The $e-$values are deduced from the elliptical
isophote fit to the cluster light described in the text.
The cluster has a small and nearly uniform eccentricity 
$\langle e \rangle \simeq 0.06$.
\label{fig5}}
\end{figure}

Cluster C44 is sufficiently extended on the PC1 image that we can 
make useful estimates of its structural parameters.  The smooth profile of the
cluster light generated from our {\it stsdas.ellipse} and {\it .bmodel}
fits is shown in Figure~\ref{fig4} .  Both the $V$ and $I$ surface intensity
plots indicate that the core of the cluster is 
significantly more extended than the PSF profile, and
at large radii, the light can be 
traced out past $r \sim 2''$ (40 px).  The profile shape is 
also nearly circular and fairly uniform with radius, 
as shown in Figure~\ref{fig5}; the eccentricity of
the model fit goes through modest fluctuations 
about its mean value $\langle e \rangle \simeq 0.06$.

Comparison with representative \cite{kin66} model profile
curves, also shown in Figure~\ref{fig4}, indicates that the cluster 
is adequately matched by models with central potential
parameters in the range $W_0 = 7 \pm 1$.
A {\it very rough} estimate of the core radius of
the cluster from the model fits is 
$r_c \simeq 0\farcs08 \simeq 1.5$ pc, after subtraction in 
quadrature of the PSF ``core'' $r_{c,PSF} \simeq 0\farcs04$.
The half-mass radius, which is larger and thus more
well determined, is at $r_h \simeq 0\farcs27 \simeq 5$ pc, and the
central concentration index of the fitted curve is
$c =$ log$(r_t/r_c) = 1.6 \pm 0.3$.
Its estimated tidal radius is then $r_t \sim 3'' \sim 60$ pc.  These
values all lie comfortably within the 
typical range for globular clusters in the outer
halo of the Milky Way, which have more extended cores
and larger tidal radii than those in the inner bulge 
(\eg, \cite{tra95}; \cite{vdb91}).

In summary, this study clearly favors the
important ``null hypothesis'' that N5128-C44 is 
an old, metal-poor globular cluster closely resembling the familiar
ones in the Milky Way in every way we can test it:  
metallicity, integrated color, color-magnitude diagram, 
and King-like structural parameters.
This is the first time that direct stellar photometry has been used
to test this hypothesis in any globular
cluster in a giant E galaxy.  Our results support the contention
that we are genuinely dealing with the same fundamental type of
object when we perform comparisons between old-halo globular cluster
populations in gE galaxies and those in the Local Group.
We suggest that the photometric analysis
of globular clusters in other galaxies well beyond the Local
Group is within reach, and will become routine with the deployment
of more advanced imaging tools.

\acknowledgments

We are grateful for technical advice from Pat Durrell and
Dean McLaughlin.
This work was supported financially by the Natural Sciences and 
Engineering Research Council of Canada, through operating grants
to G.L.H.H. and W.E.H.

\clearpage

%
%

\clearpage

\clearpage

\clearpage

\clearpage

\clearpage


\begin{thebibliography}{}

\bibitem[Christian \& Heasley 1991]{chr91} Christian, C.~A.,
\& Heasley, J.~N. 1991, \aj, 101, 848

\bibitem[Cohen \etal\ 1997]{coh97} Cohen, J.~G., Blakeslee, J.~P.,
\& Ryzhov, A. 1997, \apj, 496, 808

\bibitem[Couture et al.\ 1995]{cou95} Couture, J., Racine, R.,
Harris, W.~E., \& Holland, S. 1995, \aj, 109, 2050

\bibitem[Da Costa \& Armandroff 1990]{dac90} Da Costa, G.,
\& Armandroff, T. 1990, \aj, 100, 162

\bibitem[Ebneter \& Balick 1983]{ebn83} Ebneter, K., \& Balick, B. 1983,
\pasp, 95, 675

\bibitem[Fernley \etal\ 1998]{fer98} Fernley, J. \etal\ 1998,
\aap, 330, 515

\bibitem[Fusi Pecci \etal\ 1996]{fus96} Fusi Pecci, F.,
\etal\ 1996, \aj, 112, 1461

\bibitem[Graham 1979]{gra79} Graham, J.~A. 1979, \apj, 232, 60

\bibitem[Guarnieri et al.\ 1998]{gua98} Guarnieri, A.,
Clementini, G., Valentini, G., Castro-Tirado, A., Gorosabel, J.,
\& Pedrosa, A. 1998, \aap, 331, 70

\bibitem[G.~Harris \etal\ 1984]{har84} Harris, G.~L.~H., Hesser, J.~E.,
Harris, H.~C., \& Curry, P.~J. 1984, \apj, 287, 175

\bibitem[G.~Harris et al.\ 1992]{hghh92} Harris, G.~L.~H.,
Geisler, D., Harris, H.~C., \& Hesser, J.~E. 1992, \aj, 104, 613 (HGHH)

\bibitem[G.~Harris \etal\ 1998]{har98} Harris, G.~L.~H., Harris, W.~E.,
\& Poole, G.~B. 1998, in preparation

\bibitem[H.~Harris \etal\ 1988]{har88} Harris, H.~C.,
Harris, G.~L.~H., \& Hesser, J.~E. 1988, in Globular Cluster Systems
in Galaxies, IAU Symposium No.~126, ed.~J.~E.~Grindlay \& A.~G.~D.~Philip
(Dordrecht:  Kluwer), 205

\bibitem[W.~Harris 1991]{har91} Harris, W.~E., \araa, 29, 543 

\bibitem[Heasley \etal\ 1988]{hea88} Heasley, J.~N., Christian, C.~A.,
Friel, E.~D., \& Janes, K.~A. 1988, \aj, 96, 1312

\bibitem[Hesser \etal\ 1986]{hes86} Hesser, J.~E., 
Harris, H.~C., \& Harris, G.~L.~H. 1986, \apjl, 303, L51

\bibitem[Hesser \etal\ 1984]{hes84} Hesser, J.~E., Harris, H.~C.,
van den Bergh, S., \& Harris, G.~L.~H. 1984, \apj, 276, 491

\bibitem[Holtzman \etal\ (1995)]{hol95} Holtzman, J.~A. 1995,
\pasp, 107, 1065

\bibitem[Hui \etal\ 1993]{hui93} Hui, X., Ford, H.~C.,
Ciardullo, R., \& Jacoby, G.~H. 1993, \apj, 414, 463

\bibitem[Hui \etal\ 1995]{hui95} Hui, X., Ford, H.~C.,
Freeman, K.~C., \& Dopita, M.~A. 1995, \apj, 449, 592

\bibitem[Jablonka \etal\ 1996]{jab96} Jablonka, P., Bica, E.,
Pelat, D., \& Alloin, D. 1996, \aap, 307, 385

\bibitem[King (1966)]{kin66} King, I.~R. 1966, \aj, 71, 64

\bibitem[Rich \etal\ 1996]{rich96} Rich, R.~M., Mighell, K.~J.,
Freedman, W.~L., \& Neill, J.~D. 1996, \aj, 111, 768

\bibitem[Soria \etal\ 1996]{sor96} Soria, R. \etal\ 1996,
\apj, 465, 79

\bibitem[Stetson 1992]{ste92} Stetson, P.B. 1992, 
in Astronomical Data Analysis Software and Systems I, 
ASP Conf.Ser. Vol.8, edited by G.H.Jacoby (ASP, San Francisco), p.~289

\bibitem[Storchi-Bergmann \etal\ 1997]{sto97} Storchi-Bergmann, T.,
Bica, E., \& Kinney, A.~L. 1997, \mnras, 290, 231

\bibitem[Tonry \& Schechter 1990]{ton90} Tonry, J.~L., \& 
Schechter, P.~L. 1990, \aj, 100, 1794

\bibitem[Trager \etal\ 1995]{tra95} Trager, S.~C., King, I.~R.,
\& Djorgovski, S. 1995, \aj, 109, 218

\bibitem[van den Bergh 1995]{vdb95} van den Bergh, S. 1995,
\apj, 446, 39

\bibitem[van den Bergh \etal\ 1991]{vdb91} van den Bergh, S.,
Morbey, C., \& Pazder, J. 1991, \apj, 375, 594


\end{thebibliography}
\end{document}